\documentclass[12pt,preprint]{aastex}

\newcommand{\avgm}{\langle M \rangle}

\shorttitle{X-ray Microlensing in PG 1115+080}
\shortauthors{Morgan et al.}

\begin{document}

\title{X-Ray and Optical Microlensing in the Lensed Quasar PG~1115+080\altaffilmark{1}}

\author{Christopher W. Morgan\altaffilmark{2,3}, Christopher.S. Kochanek\altaffilmark{3,4}, Xinyu Dai\altaffilmark{3,4}, 
Nicholas D. Morgan\altaffilmark{3} and Emilio E. Falco\altaffilmark{5}}

\altaffiltext{1}{Based on
observations obtained with the Small and Moderate Aperture Research Telescope
System (SMARTS) 1.3m, which is operated by the SMARTS Consortium,
the Apache Point Observatory 3.5-meter telescope, which is
owned and operated by the Astrophysical Research Consortium, the WIYN Observatory
which is owned and operated by the University of Wisconsin, Indiana University,
Yale University and the National Optical Astronomy Observatories (NOAO), the
6.5m Magellan Baade telescope, which is a collaboration between the observatories
of the Carnegie Institution of Washington (OCIW), University of Arizona,
Harvard University, University of Michigan, and Massachusetts Institute
of Technology, and observations made with the NASA/ESA Hubble Space Telescope
for program HST-GO-9744 of the Space Telescope Science Institute,
which is operated by the Association of Universities for Research
in Astronomy, Inc., under NASA contract NAS 5-26555.}
\altaffiltext{2}{Department of Physics, United States Naval Academy, 572C Holloway Road,
Annapolis, MD 21402}
\altaffiltext{3}{Department of Astronomy, The Ohio State University, 140 West 18th Avenue, 
Columbus, OH 43210-1173}
\altaffiltext{4}{Center for Cosmology and AstroParticle Physics, The Ohio State University}
\altaffiltext{5}{Harvard-Smithsonian Center for Astrophysics, 60 Garden Street, Cambridge, MA, 02138} 

\clearpage

\begin{abstract}
We analyzed the microlensing of the X-ray and optical emission of the
lensed quasar PG 1115+080. We find that 
the effective radius of the X-ray emission is $1.3^{+1.1}_{-0.5}$~dex 
smaller than that of the optical emission.  
Viewed as a thin disk observed at inclination angle $i$, the optical accretion disk has a 
scale length, defined by the point where the disk temperature matches the rest frame energy of the 
monitoring band ($kT=hc/\lambda_{rest}$ with $\lambda_{rest}=0.3\mu$m), 
of $\log[(r_{s,opt}/ {\rm cm}) \sqrt{\cos(i) / 0.5}] = 16.6 \pm 0.4$ .  
The X-ray emission region (1.4-21.8 keV in the rest frame) has an effective half-light radius of 
$\log[r_{1/2,X}/{\rm cm}] = 15.6^{+0.6}_{-0.9}$.  
Given an estimated black hole mass of $1.2 \times 10^9 \, {\rm M_\sun}$,
corresponding to a gravitational radius of $\log [r_{g}/{\rm cm}] = 14.3$, 
the X-ray emission is generated near the inner edge of the disk while the optical emission comes 
from scales slightly larger than those expected for an Eddington-limited thin disk.  
We find a weak trend supporting models with low stellar mass fractions
near the lensed images, in mild contradiction to inferences from the stellar velocity 
dispersion and the time delays. 
\end{abstract}


\section{Introduction}
\label{sec:introduction}
When \citet{Blaes2007} recently reviewed 
the state of accretion disk physics, he found that one of the most 
glaring problems in even the most sophisticated accretion disk models 
\citep[e.g.][]{Hubeny2000,Hubeny2001,Hirose2006}
is their failure to support a hot corona or to produce X-rays at all.
While there are models for producing the X-rays,
\citep[e.g.][]{Haardt1991,Haardt1993,Maraschi2003,Hirose2004,Nayakshin2004}, they do so
on very different physical scales relative to the gravitational radius $r_g=GM_{BH}/c^2$ of the black hole.
For example, the magnetohydrodynamic (MHD) simulations of \citet{Hawley2002} predict the dragging of hot
ionized gas from a jet across the surface of a cooler accretion disk resulting in bremsstrahlung.  In this model, 
much of the emission comes from an inner torus with radius $r \lesssim 20 r_{g}$, 
but the continuum emission region extends to very large radii ($r \approx 200 r_{g}$). 
On much smaller scales, the model of \citet{Hirose2004} suggests
a relativistic MHD accretion disk model in the Kerr metric whose 
inner torus ($r \lesssim 10 r_{g}$) supports a large current density capable of emitting a moderate X-ray flux. 
The disk-corona model of \citet[][see also Merloni 2003]{Haardt1991} produces
X-rays via inverse Compton scattering in a corona which extends over much of the
optical/UV accretion disk, while the ``lamp-post'' \citep{Martocchia2002} and
``aborted jet'' \citep{Ghisellini2004} models predict a significantly smaller emission structure 
($r \lesssim 3.0 r_{g}$).

Given their small angular size, few traditional observational 
constraints can be placed on the size of quasar X-ray continuum emission regions apart from 
simple and often inconclusive variability timescale arguments \citep[e.g.][]{Vaughan2003}.
\citet{Fabian1995} demonstrated that the broad Fe K$\alpha$ X-ray emission line in Seyfert 1
spectra is probably emitted from the region immediately surrounding the black hole. 
The width and variability of Fe K$\alpha$ emission has now been measured in a number of systems to 
study the innermost regions of those accretion disks
\citep[e.g.][]{Iwasawa1999,Lee1999,Fabian2002,Iwasawa2004}. Motivated by this work, 
\citet{Young2000,Ballantyne2005} and others have proposed the use of {\it Constellation-X}
to measure the size of the Fe K$\alpha$ X-ray reflection region by reverberation mapping. 
Fortunately, gravitationally lensed quasars can be studied on these scales at
all wavelengths because the quasar is microlensed by the stars in
the lens galaxy. The Einstein radius $R_E$ of the stars is comparable
to the expected near-IR sizes of accretion disks, so most disk
emission will be significantly microlensed with the amplitude
of the variability increasing rapidly for source components that
are small compared to $R_E$ due to the presence of caustic curves
on which the microlensing magnification diverges \citep[see the review by][]{Wambsganss2006}.

More generally, microlensing variability is a function
of the relative tangential velocity $v_e$ between source,
lens and observer, the macroscopic lensing properties of the lens galaxy
(the convergence $\kappa$, the stellar surface density fraction $\kappa_{*}/\kappa$ and 
the shear $\gamma$) and the relative sizes of the source and the 
source plane projection of the Einstein radius $R_E$ of an average mass star $\avgm$ in the lens galaxy. 
Since the size of the X-ray emitting region is expected to be much smaller than the optical
accretion disk, we expect that the effects of microlensing will be more pronounced at X-ray 
wavelengths than in the optical \citep[e.g.][]{Jovanovic2008}. This effect has now been observed in many
lensed quasars 
\citep[e.g.][]{Morgan2001,Chartas2002,Dai2003,Blackburne2006,Chartas2007,Pooley2007} and was
documented specifically in PG 1115+080 by \citet{Pooley2006}.

The quadruply-lensed quasar PG 1115+080 was discovered over 25 years ago \citep{Weymann1980}. Since then, it has
been the subject of a large number of investigations at multiple wavelengths. In particular, the closely separated A1 and A2
images bracket a critical line so we expect their flux ratio to be approximately unity, 
but \citet{Impey1998} and others have measured an anomalously low flux ratio in the
optical and NIR (e.g. $A2/A1=0.64 \pm 0.02$ in the $H$-band).  \citet{Chiba2005} showed that
the $A2/A1$ flux ratio returns to nearly unity in the mid-IR ($A2/A1 = 0.93 \pm 0.06$~at $11.7 \micron$), 
demonstrating that stellar microlensing is the likely cause of the anomaly rather than
millilensing \citep[e.g.][]{KochanekandDalal2004}. 

Recently, \citet{Pooley2006}
conducted a study of the system's anomalous X-ray flux ratios as measured in two {\it Chandra
X-Ray Observatory} ({\it Chandra}) observations. \citet{Pooley2006} demonstrated that microlensing
is the likely cause of the X-ray flux ratio anomaly in PG 1115+080 and qualitatively argued that 
its X-ray continuum emission region must be significantly smaller than its optical accretion
disk.  In this paper, we combine these 2 epochs of X-ray data from {\it Chandra} with our optical
monitoring data to make simultaneous measurements of the system's optical and X-ray continuum
emission regions using the Monte Carlo microlensing analysis technique of \citet{Kochanek2004}.

In \S~\ref{sec:observations}, we describe our optical monitoring data and the
X-ray flux measurements. In \S~\ref{sec:microlensing}, we review our 
microlensing analysis technique and describe its application to  
PG 1115+080. In \S~\ref{sec:results} we present
the results of our calculations and discuss their implications
for the sizes of the quasar emission regions and the stellar content of the lens galaxy. 
We assume a flat cosmology with $\Omega_0 = 0.3$, $\Lambda_0 = 0.7$ and
$H_0=70 \: {\rm km \: s^{-1} \: Mpc^{-1}}$.

\section{{\it Hubble Space Telescope} Observations, {\it Chandra} and Optical Monitoring Data}
\label{sec:observations}

We observed PG 1115+080 in the $V$- (F555W), $I$- (F814W) and $H$- (F160W) bands using
the {\it Hubble Space Telescope} ({\it HST}) for the CfA-Arizona Space Telescope Survey 
\citep[CASTLES\footnote{http://cfa.harvard.edu/castles/},][]{CASTLES}. 
The $V$- and $I$-band images were taken using the Wide-Field Planetary Camera 2 (WFPC2). 
The $H-$band images, originally reported in \citet{Yoo2005},
were taken using the Near-Infrared Camera and Multi-Object
Spectrograph (NICMOS). We made photometric and astrometric fits to the {\it HST} imagery 
with {\it imfitfits} \citep{Lehar2000}, using a de Vaucouleurs model for the lens galaxy, 
an exponential disk model for the quasar host galaxy and point sources for the quasar images.
The astrometric fits are consistent with those of \citet{Impey1998}. 
Our {\it HST} astrometry and photometry are presented in Table~\ref{tab:hst}. 

We monitored PG 1115+080 in the $R$-band over multiple seasons with the SMARTS 1.3m
telescope using the ANDICAM optical/infrared camera 
\citep{Depoy2003}\footnote{http://www.astronomy.ohio-state.edu/ANDICAM/},
the Wisconsin-Yale-Indiana (WIYN) observatory
using the WIYN Tip--Tilt Module (WTTM) \footnote{http://www.wiyn.org/wttm/WTTM\_manual.html},
the 2.4m telescope at the MDM Observatory using the MDM 
Eight-K\footnote{http://www.astro.columbia.edu/~arlin/MDM8K/}, Echelle and 
RETROCAM\footnote{http://www.astronomy.ohio-state.edu/MDM/RETROCAM} \citep{Morgan2005}
imagers and the 6.5m Magellan Baade telescope using IMACS \citep{Bigelow1999}.
A detailed discussion of our lensed quasar monitoring data reduction pipeline can be found in 
\citet{Kochanek2006}, but we briefly summarize our technique here.
We hold the lens astrometry fixed to the {\it HST} $H$-band measurements.
We treat each quasar image as a point source and model the point-spread function
with three nested, elliptical Gaussian profiles.  
We measure the flux of each image by comparison to the flux of 5 reference stars
in the field.
We assume that the lens galaxy flux remains constant and fix its value to the flux 
found by minimizing the residuals in a fit to the complete set of measurements from each instrument.
We supplemented our optical lightcurves with $V$-band data published by \citet{Schechter1997}.
The \citet{Schechter1997} data set does not overlap with any 
of our new monitoring data, so we were unable to correct it for the wavelength difference between monitoring bands.
Fortunately, this difference is 
small enough to have little effect on the results given the expected $\lambda^{4/3}$ scaling
of the optical accretion disk size \citep{Shakura1973} and our measurement uncertainties.
We applied magnitude offsets \citep[e.g.][]{Ofek2003} to the monitoring data from the other observatories to 
match the $R$-band measurements from SMARTS.  

PG 1115+080 is a particularly challenging system to monitor because the A1 and A2 images are separated by
a mere $0\farcs48$. The seeing in our ground-based observations was rarely better than $1\farcs0$,
so we were forced to sum the flux from images A1 and A2. We refer to this summed
lightcurve as $A12 = A1 + A2$.  As documented by \citet{Pooley2006}, the strongest
effect of optical microlensing appears in the A1/A2 flux ratio,   
so we supplemented our lightcurves
with 7 epochs of data from the literature in which the A1 and A2 images are clearly resolved
\citep{Schechter1997,Courbin1997,Impey1998,Pooley2006}.
We present our optical monitoring data in Table~\ref{tab:lightcurves}.

We complement our optical lightcurves with X-ray fluxes from the two epochs 
of  $0.5 - 8$~keV {\it Chandra} imagery published by \citet{Pooley2006}, 
although here we used the refined flux measurements presented in \citet{Pooley2007}.
The details of the X-ray data reduction and flux ratio calculations are found in those papers. 

\section{Microlensing Models}
\label{sec:microlensing}

Microlensing statistics are strongly influenced by the presence of smoothly
distributed dark matter \citep{Schechter2002}, typically parameterized as
the ratio of the stellar surface density to the
total surface density $\kappa_{*}/\kappa$. 
We considered a range of possible stellar mass fractions in our calculations.
We used the {\it GRAVLENS} software package \citep{GRAVLENS} to generate a series of ten models 
that match the {\it HST} astrometry and reproduce the mid-infrared ($11.7 \: \micron$) 
flux ratios from \citet{Chiba2005}. 
Each model consists of concentric de Vaucouleurs and NFW \citep{Navarro1996} profiles,
and we vary the mass in the de Vaucouleurs component over the range
$0.1 \leq f_{M/L} \leq 1.0$ in steps of $\Delta f_{M/L}=0.1$, 
where $f_{M/L}=1.0$ represents a constant mass-to-light ratio (de
Vaucouleurs) model with no dark matter halo. Table~\ref{tab:models} summarizes the microlensing parameters
as a function of $f_{M/L}$.  \citet{Treu2002} found that the best fit to the system's large
stellar velocity dispersion ($\sigma_*=281 \pm 25 \; {\rm km \: s^{-1}}$, \citealt{Tonry1998}) 
is provided by a steep mass profile $\rho \propto r^{-2.35}$, implying a large stellar mass component, 
and for $H_0 = 72 \: {\rm km \: s^{-1} \: Mpc^{-1}}$, the best fit 
to the \citet{Schechter1997} time delays is provided by the $f_{M/L}=0.8$ model. 

We generated a set of microlensing magnification patterns at each image location 
for each of the 10 macroscopic mass models
using a variant of the ray-shooting method (\citealt{Schneider1992}, see \citealt{Kochanek2004}
for the details of our technique).
The patterns are $8192 \times 8192$ images of the source-plane projection of the 
magnification patterns from an ensemble of typical lens galaxy 
stars at each image location.  We approximated the 
Galactic stellar mass function of \citet{Gould2000} as a power law, assuming $dN(M)/dM \propto M^{-1.3}$
with a dynamic range in mass of a factor of 50.  The mean stellar mass in the lens galaxy $\avgm$
is initially unknown, so magnification patterns are produced in units of the Einstein radius with
an outer scale of $20 \, R_E$. For PG 1115+080, the Einstein radius is 
$R_E = 6.6 \times 10^{16} \langle M / M_\sun \rangle^{1/2}$~cm. 
To convert to physical units, all results are eventually scaled by some factor of $\langle M/M_\sun \rangle$.
Henceforth, quantities in Einstein units will be given the ``hat'' accent to distinguish
them from quantities in physical units. So the physical source size $r_s$ is related to the scaled source
size $\hat{r}_s$ by $r_s = \hat{r}_s \langle M/M_\sun \rangle^{1/2}$, and the physical effective velocity $v_e$
is related to the scaled velocity $\hat{v}_e$ by $v_e = \hat{v}_e \langle M/M_\sun \rangle^{1/2}$.

In order to eliminate the quasar's intrinsic variability, 
we shifted the optical light curves by the measured time delays \citep{Schechter1997}, 
so that any remaining variability in the
flux ratios must be attributed to microlensing. It is impossible to offset the
sparse X-ray flux measurements by the time delays,   
so we assume that X-ray flux ratios can be treated as simultaneous in a statistical sense.  
The time delay between the A1 and A2 images is less than one day, so there was no need to
apply a time delay correction to the 7 epochs of individually resolved A1 and A2 data. 
 
As described in detail by \citet{Kochanek2004} \citep[see also][]{Morgan2007,Poindexter2007}, 
our Monte Carlo microlensing analysis searches for trajectories across the magnification pattern
that fit the observed light curves.
We used a thin accretion disk surface brightness profile for the source model \citep{Shakura1973} with
\begin{equation}
I(R) \propto \left\{ \exp \left[ \left( R/r_s \right)^{3/4} \right] - 1 \right\}^{-1},
\label{eqn:thindisk}  
\end{equation}
where the scale radius $r_s$ is the radius at which the disk temperature matches the 
rest-fame wavelength of our monitoring band, $kT = h c (1+z_s)/\lambda_{obs}$. We
neglect the central hole in the emission profile, the effect of which is negligible at optical
wavelengths. Microlensing primarily depends on the projected area of the
source while the true scale lengths also depend on the shape
of the source and its inclination.  We will refer to a radius where we 
have ignored the shape and inclination of the source as an ``effective'' 
radius that defines a projected area $\pi r_{eff}^2$. For a thin disk, 
the effective radius is related to the source scale length by 
$r_{eff}^2 = r_s^2 \cos i$ where $i$ is the inclination angle.  The X-ray emission presumably
has a different emission profile and shape.  Fortunately, \citet{Mortonson2005} 
demonstrated that the half-light radius measured with microlensing is essentially independent
of the surface brightness profile, so we will
characterize the X-ray emission by the effective half-light radius.  For our thin disk model, the half-light 
radius is related to the disk scale length by $R_{1/2}=2.44 r_s$.  In summary, to compare the
sizes of the optical and X-ray emitting regions we will use the ratio of the 
effective radii $r_{opt}/r_X$, to characterize the optical emission we will use the 
thin disk scale length $r_{s,opt}$ and an inclination angle $\cos i$, and for the X-ray emission we
will use the effective (i.e. no shape corrections) half-light radius $r_{1/2,X}$.

\begin{figure}
\epsscale{1.0}
\plotone{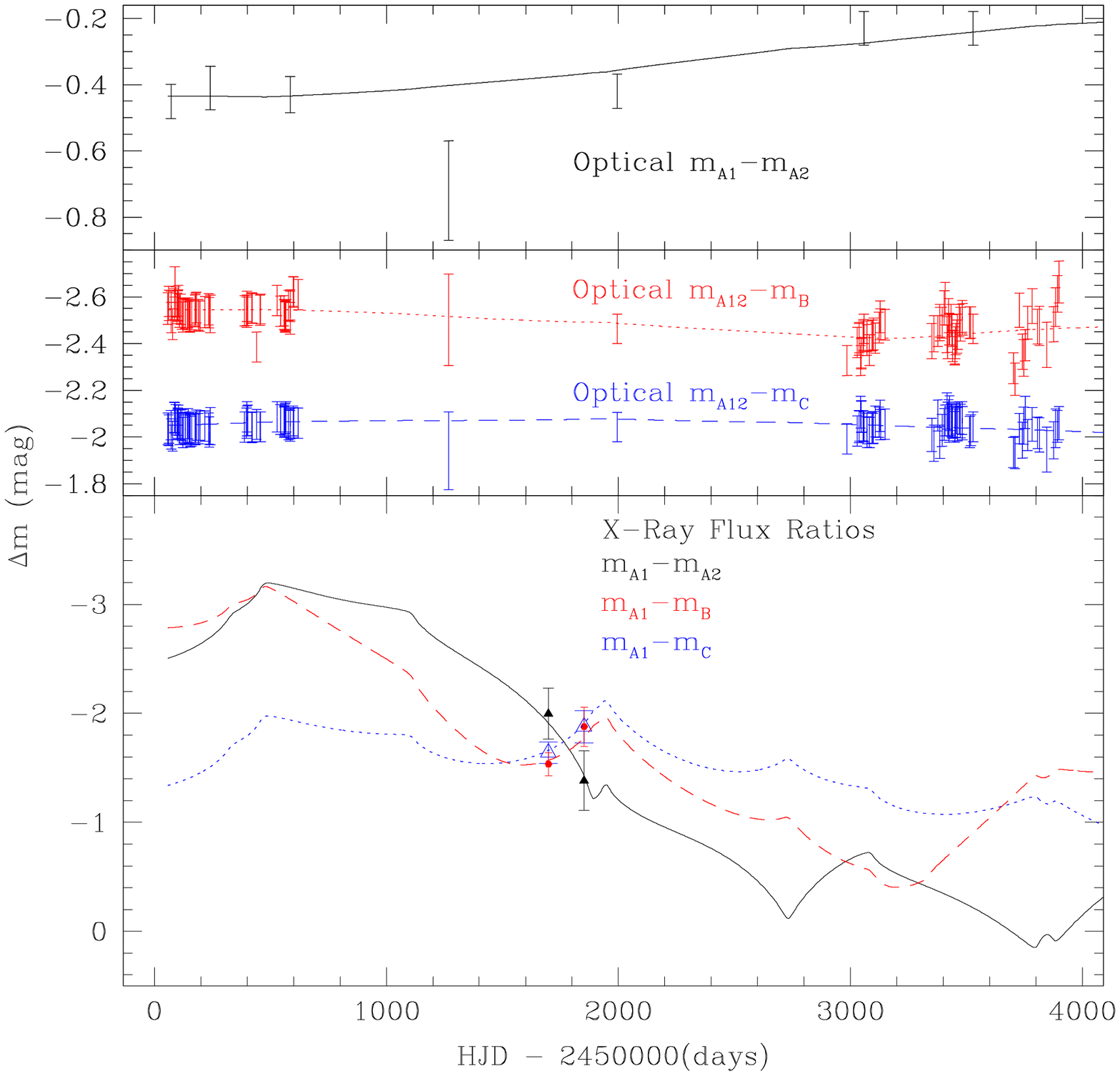}
\caption{Examples of good fits to the observed flux ratios. Top panel: We fit the 7 epochs 
of data with a resolved A1/A2 flux ratio individually.
The error bars are the data, and the black curve is the fit. 
Middle panel: Dotted red and dashed blue curves are best fits to the $A12/B$ and $A12/C$
flux ratios, respectively. Data are plotted with error bars in the same color scheme. The 
$A12/B$ and $A12/C$ flux ratios varied little over the last decade.
Bottom panel: The observed $A1/A2$, $A1/B$ and $A1/C$ X-ray flux ratios are 
plotted using solid black triangles, solid red circles and open blue triangles, 
respectively.  The best fits to the observed $A1/A2$, $A1/B$, and $A1/C$
flux ratios are plotted using solid black, dotted red and dashed blue
curves, respectively. 
\label{fig:bestfits1}}
\end{figure}

\begin{figure}
\epsscale{1.0}
\plotone{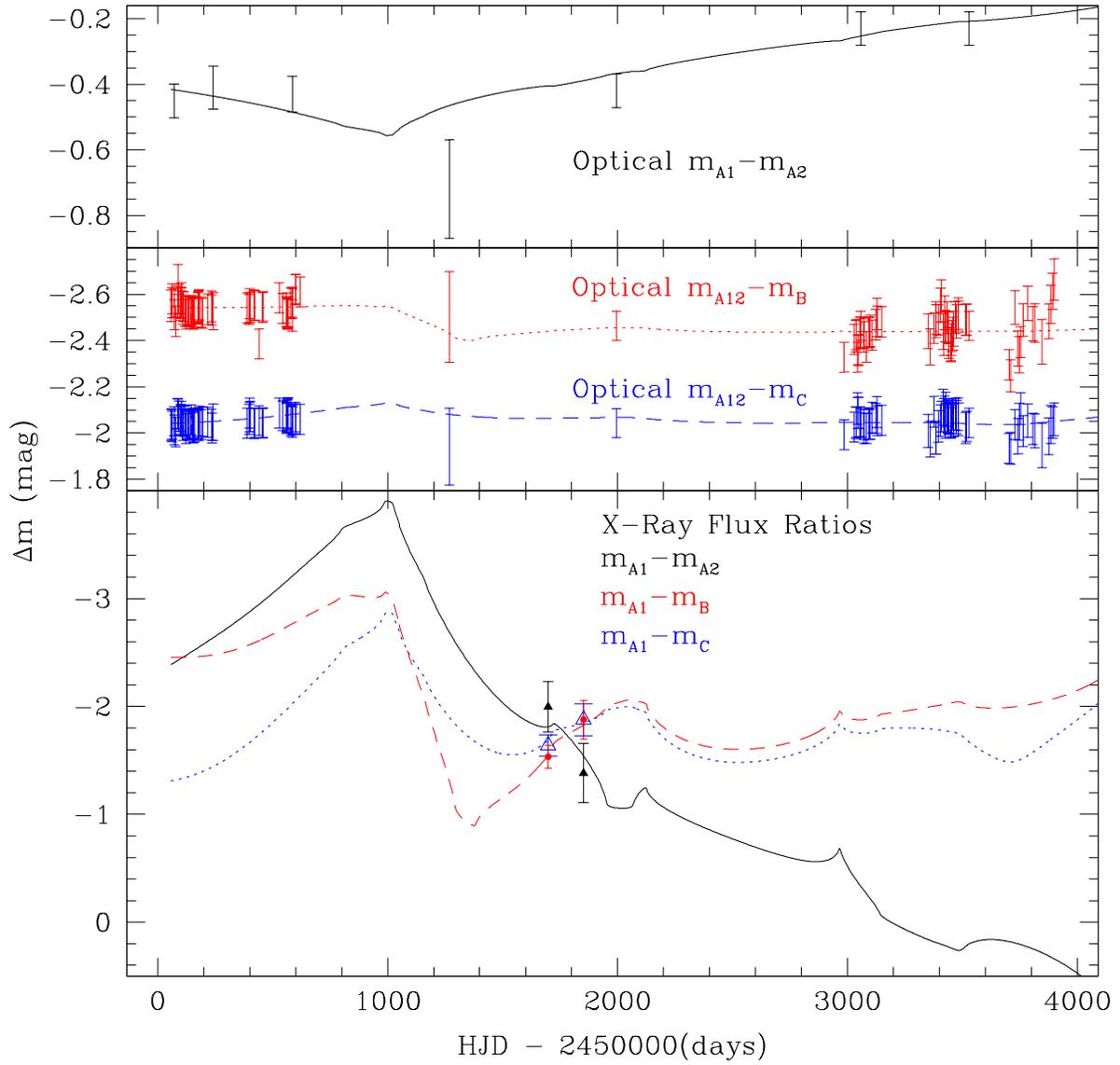}
\caption{A second solution plotted as in Fig.~\ref{fig:bestfits1}.   
\label{fig:bestfits2}}
\end{figure}

We generated 8 trial magnification patterns for each of the 10 macroscopic mass models.  For each
trial and model we produced 50,000 trial light curves for a $16 \times 21$ grid of X-ray and optical 
source sizes.  These source sizes $\hat{r}_s$ are scaled sizes that depend on the microlens 
mass $r_s = \hat{r}_s \langle M/ M_\sun \rangle^{1/2}$. We used logarithmic grids spanning the region 
producing acceptable fits with a grid spacing of 0.2~dex.
In total there were $4\times10^6$ trial light curves for each combination of X-ray and optical source 
sizes. When assessing the quality of our fits to the observed flux ratios, we allowed for 
only $0.1$~mag of systematic uncertainty in the flux ratios of the macro models 
because the mid-IR flux ratios of \citet{Chiba2005} are a close approximation 
to the intrinsic flux ratios in this system.  
In selecting trial light curves we gave equal statistical weight to the optical data
where A1 and A2 could not be separately measured, optical data where A1 and A2 could be separately 
measured, and the X-ray data so that we would isolate trials with reasonable fits to all three
classes of data.  The final goodness of fit was evaluated with a $\chi^2$ fit to the light curves 
where all data have their true statistical weights and we discard all fits with 
$\chi^2/N_{dof} > 4.0$ as they make no 
significant contribution to the final Bayesian integrals \citep[see][]{Kochanek2004}.
Figures~\ref{fig:bestfits1} and \ref{fig:bestfits2} show two examples of good fits to the data.
The stronger flux anomalies in the X-ray data force the X-ray source to be more compact than the 
optical, leading to the much larger variability predicted for the X-ray bands relative to the optical.
  
To convert the results to physical units, we assume a prior on the mean mass of the microlenses
$\avgm$ and the transverse velocity between source, lens and observer $v_e$. 
For the mean stellar mass prior, we assume $0.1 \: {\rm M_\sun} \leq \avgm \leq 1.0 \: {\rm M_\sun}$. 
We model the effective velocity of the system with three components. We set the velocity of the observer 
$v_o = 94 \: {\rm km \: s^{-1}}$ to be the projection 
of the CMB dipole velocity \citep{Kogut1993} onto the lens plane. 
We calculate a one-dimensional stellar velocity dispersion in the lens galaxy of 
$\sigma_* = 220 \: {\rm km \: s^{-1}}$ based on the Einstein radius of its macroscopic mass
model and we assume a lens galaxy peculiar 
velocity dispersion of  $\sigma_p=235/(1+z_l) \: {\rm km \: s^{-1}} = 179 \: {\rm km \: s^{-1}}$ \citep{Kochanek2004}.
In general these two priors give similar physical size estimates if applied separately because the sizes depend only 
weakly on the mass scale $\avgm$.  Microlensing depends only on the size and velocity 
of the source in Einstein units, $\hat{r}_s = r_s/\langle M/M_\sun\rangle^{1/2}$ and 
$\hat{v}_e = v_e/\langle M/ M_\sun\rangle^{1/2}$, and a given 
level of variability can be produced either by moving a small source slowly (both $\hat{r}_s$ and 
$\hat{v}_e$ small) or a large source rapidly (both $\hat{r}_s$ and $\hat{v}_e$ large) 
with (roughly) $\hat{r}_s \propto \hat{v}_e$.  But
the mass scale implied by a given Einstein velocity scales as 
$\langle M/M_\sun \rangle = (v_e/\hat{v}_e)^2$ 
so the dependence of the physical scale on the mass essentially cancels 
given some knowledge of the physical velocity $v_e$, 
with $r_s = \hat{r}_s \langle M/M_\sun \rangle^{1/2} \propto \avgm^0$ \citep[see][]{Kochanek2004}.
For PG~1115+080 the poor temporal overlap of the optical and 
X-ray light curves means that the differences between using the priors
separately are larger than we have found for most other lenses \citep[e.g.][]{Morgan2007,Poindexter2007}, so 
we restricted our analysis to using the priors jointly.

\section{Results and Discussion}
\label{sec:results}

\begin{figure}
\epsscale{1.0}
\plotone{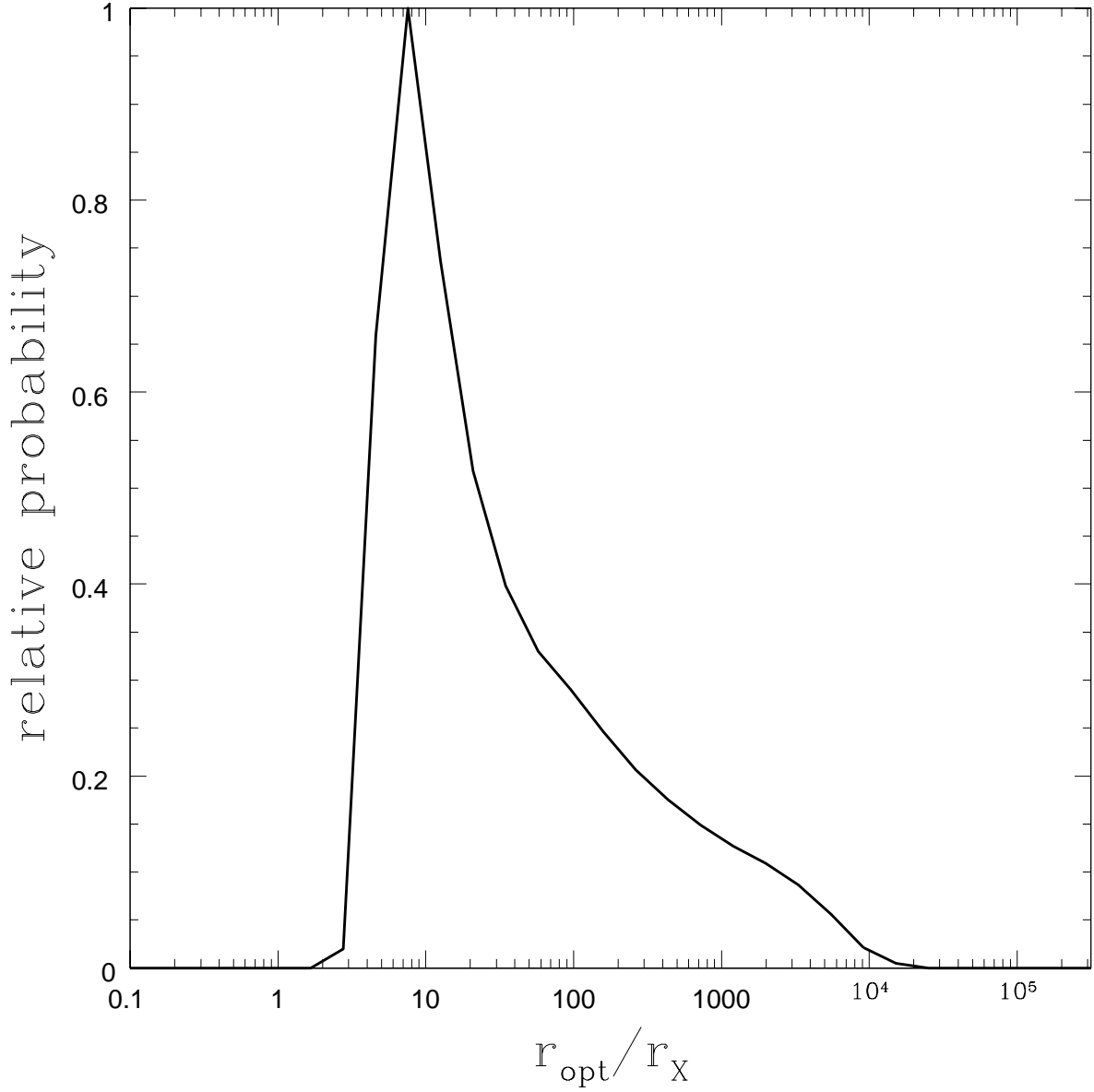}
\caption{Probability distribution for the ratio of effective radii of the
optical and X-ray sources.
\label{fig:sratio}}
\end{figure}

\begin{figure}
\epsscale{1.0}
\plotone{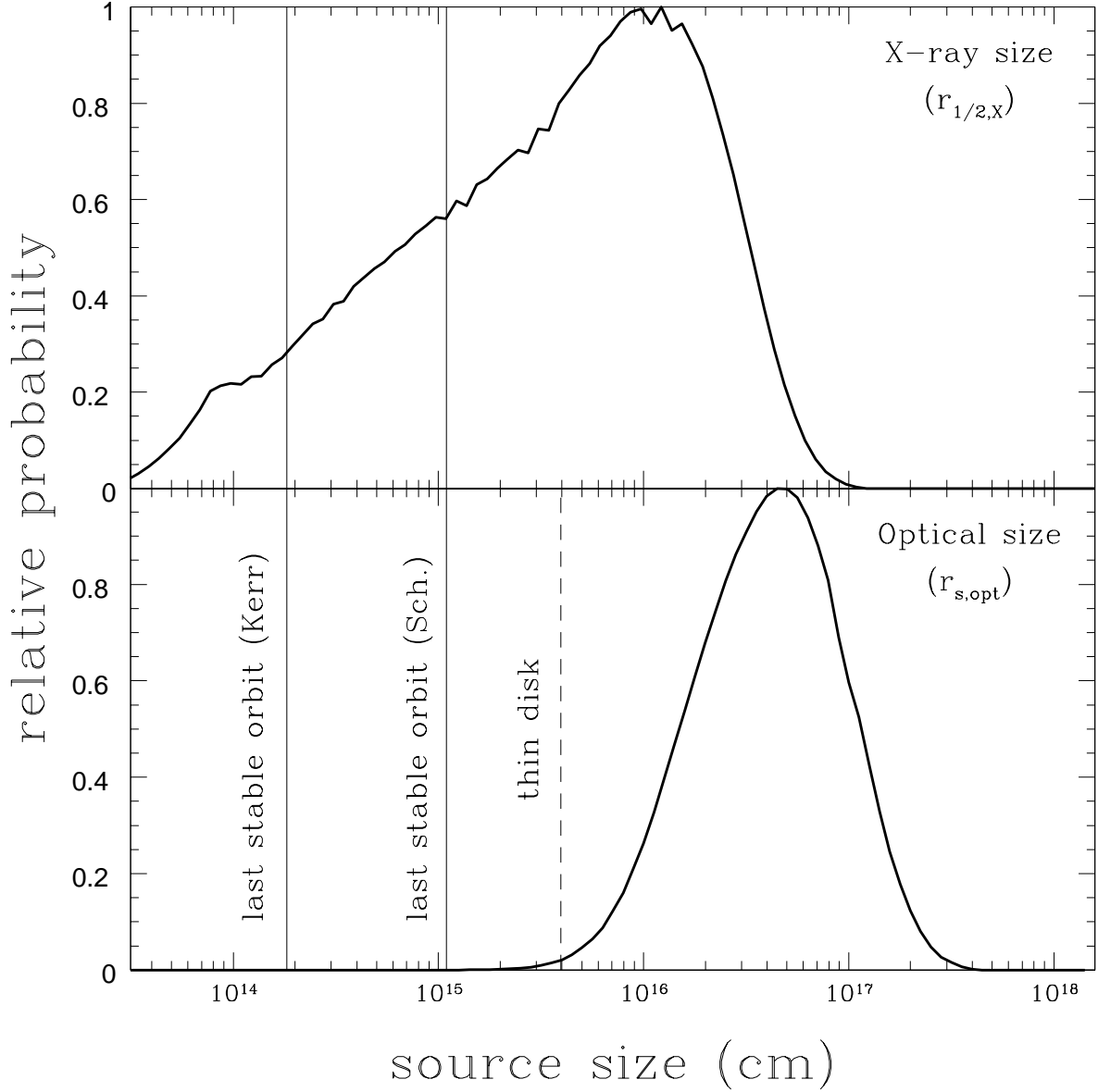}
\caption{Probability distributions for the effective X-ray half light radius $r_{1/2,X}$ (top) 
and optical thin disk scale radius $r_{s,opt}$ (bottom).  For the thin disk, we assumed an inclination of
$\cos i = 1/2$.
Given the black hole mass estimate of $1.2 \times 10^9 \: {\rm M_\sun}$ for PG 1115+080 from \citet{Peng2006},
the solid vertical lines indicate the innermost stable circular orbit $r_g=GM_{BH}/c^2$ for a maximally rotating Kerr black hole  
and the innermost stable circular orbit for a Schwarzschild black hole at $6r_{g}$.  The dashed vertical line is the 
prediction of thin disk theory for the scale radius at $0.26\micron$ for an Eddington-limited
accretion disk radiating at 10\% efficiency.  
\label{fig:rs}}
\end{figure}

Figure~\ref{fig:sratio} shows the ratio of the effective radii of the optical and X-ray sources 
where the effective radius should be viewed as the square root of the projected source area.  
The advantage of the size ratio is that it has no direct dependence on the mass 
of the microlenses (in the sense that $\hat{r}_{opt}/\hat{r}_X = r_{opt}/r_X$).  We find that 
$\log[r_{opt}/r_X] = 1.3^{+1.1}_{-0.5}$.  Figure~\ref{fig:rs} shows the estimates for the physical sizes, 
where we show the inclination corrected disk scale length for the optical source and the 
effective half-light radius for the X-ray source.  
Recall from \S~\ref{sec:microlensing} that the disk scale length is the 
point where the temperature equals the photon energy $kT=hc/\lambda_{rest}$ and that the effective 
half-light radius has no shape or inclination corrections.
Thus, at $0.3\micron$ or $T=4.8\times10^4$~K, the disk scale length 
is $\log[(r_{s,opt}/ {\rm cm}) \sqrt{\cos(i) / 0.5}] = 16.6 \pm 0.4$, and in the (rest-frame) 1.4-21.8~keV 
band the effective X-ray half light radius is $\log(r_{1/2,X}/{\rm cm}) = 15.6^{+0.6}_{-0.9}$.  

We can compare these size estimates to theoretical expectations given the estimated black hole mass of 
$1.2 \times 10^9 \, {\rm M_\sun}$ from \citet{Peng2006} based on the 
quasar luminosity and the width of the \ion{Mg}{2}~($\lambda 2798$\AA) emission line, where
black hole mass estimates using this technique have a typical uncertainties of $\sim0.3$~dex
\citep[see][]{McLure2002,Peng2006}.
Fig.~\ref{fig:rs} shows the gravitational radius $r_{g}=1.8 \times 10^{14}$~cm for this mass, which
is the innermost stable circular orbit for a maximally rotating Kerr black hole. For reference, 
we also plot the innermost stable circular orbit for a Schwarzschild black hole at $6r_{g}$.
The optical emission comes from well outside the inner edge of the disk ($\sim 220 r_{g}$), 
justifying our neglect of the inner edge of the disk in Eqn.~\ref{eqn:thindisk}.
Thin disk theory \citep{Shakura1973} predicts that the optical size for a face-on quasar 
radiating with 10\% efficiency at the Eddington limit should be $\log[r_{s,opt}/{\rm cm}]=15.6$, and
our black hole mass/accretion disk size scaling from \citet{Morgan2007} predicts a scale radius of 
$\log[(r_{s,opt}/ {\rm cm}) \sqrt{\cos(i) / 0.5}] = 15.8 \pm 0.3$. 
Our current result is marginally inconsistent with the \citet{Morgan2007} 
black hole mass - accretion disk size scaling,
but the small discrepancy could be explained by adjusting the unknown disk inclination angle. 
Our result is also somewhat larger than the theoretical thin disk size 
\citep[see][for a detailed comparison of microlensing disk size estimates in 11 systems]{Morgan2007}.  
The far more compact X-ray emission comes from a region very close to the inner disk edge, with 
an effective half-light radius of $\sim 20 r_{g}$. This result seems to favor 
models with a small or moderately-sized emission 
structures \citep[e.g.][]{Hawley2002,Martocchia2002,Ghisellini2004,Hirose2004} and disfavor 
the standard disk-corona model \citep{Haardt1991,Merloni2003}.

\begin{figure}
\epsscale{1.0}
\plotone{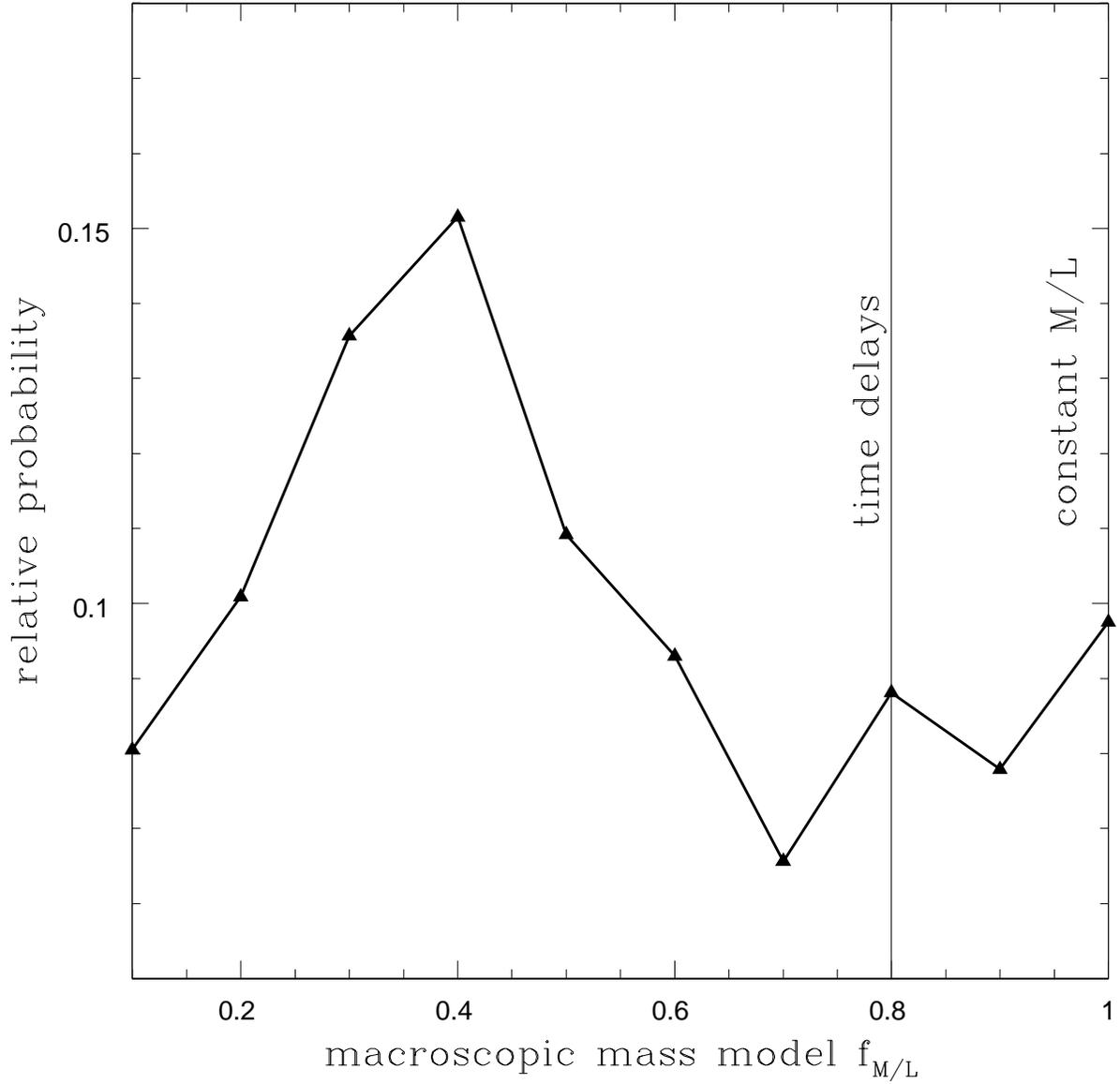}
\caption{Probability distribution for $f_{M/L}$, the fraction of the lens galaxy
mass in the constant M/L ratio (de Vaucouleurs) component. $f_{M/L}$ can
be related to the stellar surface density fraction $\kappa_*/\kappa$ 
at each image location using the data in Table~\ref{tab:models}.  The $f_{M/L}$
value implied by the time delays is plotted with a solid vertical line.
\label{fig:fml}}

\end{figure}

We also obtain some information on the structure of the lens 
galaxy, as illustrated in Figure~\ref{fig:fml} 
where we show our estimates of the stellar fraction $f_{M/L}$.  
We do not find a strong peak in the $f_{M/L}$ distribution, but we do
detect a weak trend favoring models with lower $f_{M/L}$ and
a low stellar surface density (Fig.~\ref{fig:fml}), as we would expect. 
Both the \citet{Schechter1997} time delays \citep[see][]{Kochanek2002} and the 
\citet{Tonry1998} velocity dispersion \citep[see][]{Treu2002} require mass models 
with little dark matter near the radius of the lensed images
(although see \citealt{Romanowsky1999} for examples of dynamical 
models consistent with both a significant dark matter
halo and the high velocity dispersion). 
While Figure~\ref{fig:fml} is not conclusive, it is probable that with better X-ray
light curves we will be able to measure $f_{M/L}$ and either confirm
or reject the time delay and velocity dispersion measurements.

We are expanding our analyses to include all 10 lensed quasars
with archival X-ray data \citep[see][]{Pooley2007} as well as three
systems (HE~1104--1805, RX~J1131--1231 and Q~2237+0305) where we have
obtained X-ray light curves.  For many of these systems (RX~J1131--1231, Q~2237+0305, 
WFI~J2033--4723, SDSS~0924+0219 and H~1413+117) we have optical light curves comparable to those used here
\citep[see][]{Morgan2006,Dai2007}, but for the remainder we will
have to rely on sparse, archival optical data.  The main challenge
we face is that the computational intensity of modeling the two
bands simultaneously is a significant bottleneck for completing the analyses.  
Nonetheless, we see no fundamental barriers to complementing our correlations between optical disk size and
black hole mass \citep{Morgan2007} with their X-ray equivalents. The
pattern suggested by PG~1115+080 is that the X-ray continuum emission region
tracks the inner edge of the accretion disk.  We hope to determine if this result is
universal.

\acknowledgements
CSK acknowledges support from NSF grant AST-0708082.
This research made extensive use
of a Beowulf computer cluster obtained through the Cluster Ohio
program of the Ohio Supercomputer Center. Support for program HST-GO-9744 was
provided by NASA through a grant from the Space Telescope Science Institute, which 
is operated by the Association of Universities for Research in Astronomy, Inc., 
under NASA contract NAS-5-26666.
We would like to thank P. Schechter, J. Blackburne, S. Kozlowski and D. Pooley for identifying and
helping to resolve technical issues with measurements of the A1/A2 flux ratio. 

{\it Facilities:} \facility{CTIO:2MASS (ANDICAM)}, \facility{Hiltner (RETROCAM)}, \facility{WIYN (WTTM),
\facility{HST (NICMOS, ACS)}}, \facility{CXO}.

\clearpage

\def\hm{\hphantom{-}}
\begin{deluxetable}{lccccc}
\tabletypesize{\scriptsize}
\tablewidth{0pt}
\tablecaption{HST Astrometry and Photometry of PG1115+080}
\tablehead{\colhead{Component} 
                 &\multicolumn{2}{c}{Astrometry}
                 &\multicolumn{3}{c}{Photometry}\\
                 \colhead{}
                 &\colhead{$\Delta\hbox{RA}$}
                 &\colhead{$\Delta\hbox{Dec}$}
                 &\colhead{H=F160W}
                 &\colhead{I=F814W}
                 &\colhead{V=F555W}
                 }
\startdata
A1  &$\; 1\farcs328\pm0\farcs003$ &$-2\farcs034\pm0\farcs003$ &$15.71\pm0.02$ &$16.42\pm0.02$ &$16.90\pm0.11$\\
A2  &$\; 1\farcs477\pm0\farcs004$ &$-1\farcs576\pm0\farcs003$ &$16.21\pm0.02$ &$16.85\pm0.01$ &$17.62\pm0.09$\\
B  &$-0\farcs341\pm0\farcs003$ &$-1\farcs961\pm0\farcs003$ &$17.70\pm0.02$ &$18.37\pm0.01$ &$18.95\pm0.12$\\
C  &$\equiv 0$ &$\equiv 0$ &$17.23\pm0.03$ &$17.91\pm0.02$ &$18.39\pm0.06$\\
G  &$\; 0\farcs381\pm0\farcs003$ &$-1\farcs344\pm0\farcs003$ &$16.66\pm0.04$ &$18.92\pm0.02$ &$20.74\pm0.03$\\
\enddata
\label{tab:hst}
\end{deluxetable}

\begin{deluxetable}{ccccccccccccc}
\tabletypesize{\scriptsize}
\tablewidth{0pt}
\tablecaption{PG1115+080 Lens Galaxy Mass Models}

\tablehead{\colhead{$f_{M/L}$}
                &\multicolumn{4}{c}{Convergence $\kappa$}
                &\multicolumn{4}{c}{Shear $\gamma$}
                &\multicolumn{4}{c}{$\kappa_{*}/\kappa$}\\
           	\colhead{}
                &\colhead{A1}
                &\colhead{A2}
                &\colhead{B}
                &\colhead{C}
                &\colhead{A1}
                &\colhead{A2}
                &\colhead{B}
                &\colhead{C}
                &\colhead{A1}
                &\colhead{A2}
                &\colhead{B}
                &\colhead{C}
                          }
\startdata
$0.1$ &$ 0.75 $&$ 0.76 $&$ 0.84 $&$ 0.69 $&$ 0.22 $&$ 0.27 $&$ 0.32 $&$ 0.18 $&$ 0.021 $&$ 0.023 $&$ 0.031 $&$ 0.016$ \\
$0.2$ &$ 0.69 $&$ 0.71 $&$ 0.79 $&$ 0.64 $&$ 0.27 $&$ 0.33 $&$ 0.41 $&$ 0.20 $&$ 0.046 $&$ 0.050 $&$ 0.066 $&$ 0.034$ \\
$0.3$ &$ 0.64 $&$ 0.66 $&$ 0.74 $&$ 0.59 $&$ 0.31 $&$ 0.39 $&$ 0.49 $&$ 0.23 $&$ 0.075 $&$ 0.081 $&$ 0.105 $&$ 0.055$ \\
$0.4$ &$ 0.59 $&$ 0.61 $&$ 0.69 $&$ 0.54 $&$ 0.36 $&$ 0.44 $&$ 0.58 $&$ 0.25 $&$ 0.11 $&$ 0.12 $&$ 0.15 $&$ 0.08$ \\
$0.5$ &$ 0.54 $&$ 0.56 $&$ 0.63 $&$ 0.49 $&$ 0.40 $&$ 0.50 $&$ 0.66 $&$ 0.28 $&$ 0.15 $&$ 0.16 $&$ 0.20 $&$ 0.11$ \\
$0.6$ &$ 0.49 $&$ 0.51 $&$ 0.59 $&$ 0.44 $&$ 0.44 $&$ 0.56 $&$ 0.74 $&$ 0.30 $&$ 0.20 $&$ 0.21 $&$ 0.25 $&$ 0.14$ \\
$0.7$ &$ 0.45 $&$ 0.46 $&$ 0.54 $&$ 0.39 $&$ 0.48 $&$ 0.61 $&$ 0.81 $&$ 0.32 $&$ 0.25 $&$ 0.27 $&$ 0.31 $&$ 0.18$ \\
$0.8$ &$ 0.40 $&$ 0.41 $&$ 0.49 $&$ 0.34 $&$ 0.52 $&$ 0.66 $&$ 0.89 $&$ 0.35 $&$ 0.32 $&$ 0.34 $&$ 0.39 $&$ 0.24$ \\
$0.9$ &$ 0.36 $&$ 0.38 $&$ 0.44 $&$ 0.30 $&$ 0.55 $&$ 0.71 $&$ 0.96 $&$ 0.36 $&$ 0.40 $&$ 0.42 $&$ 0.46 $&$ 0.29$ \\
$1.0$ &$ 0.31 $&$ 0.33 $&$ 0.40 $&$ 0.26 $&$ 0.60 $&$ 0.76 $&$ 1.03 $&$ 0.39 $&$ 0.51 $&$ 0.54 $&$ 0.57 $&$ 0.39$ \\

\enddata
\tablecomments{Convergence $\kappa$, shear $\gamma$ and the fraction of the total surface density 
composed of stars $\kappa_{*}/\kappa$ at each image location for the series of 
macroscopic mass models where $f_{M/L}=1.0$ corresponds to a constant mass-to-light
ratio model for the lens galaxy.}
\label{tab:models}
\end{deluxetable}

\def\hm{\hphantom{-}}
\begin{deluxetable}{ccccccc}
\tabletypesize{\scriptsize}
\tablecaption{PG1115+080 Optical Light Curves}
\tablewidth{0pt}
\tablehead{ HJD &\multicolumn{1}{c}{$\chi^2/N_{dof}$}
                &\multicolumn{1}{c}{Images A1+A2} 
                &\multicolumn{1}{c}{Image B} &\multicolumn{1}{c}{Image C}
                &\multicolumn{1}{c}{$\langle\hbox{Stars}\rangle$}
                &\multicolumn{1}{c}{Source} 
              }
\startdata
$2994.927$ &$  1.00$ &$-1.095\pm 0.006$ &$ 1.192\pm 0.010$ &$ 0.896\pm 0.008$ &$\hm 0.019\pm 0.004$ &  MDM \\ 
$3038.714$ &$  1.23$ &$-1.071\pm 0.008$ &$ 1.319\pm 0.012$ &$ 0.971\pm 0.010$ &$-0.014\pm 0.005$ &SMARTS \\ 
$3045.790$ &$  1.18$ &$-1.078\pm 0.014$ &$ 1.323\pm 0.018$ &$ 1.064\pm 0.024$ &$-0.046\pm 0.005$ &SMARTS \\ 
$3047.003$ &$  1.69$ &$-1.078\pm 0.004$ &$ 1.336\pm 0.007$ &$ 0.938\pm 0.007$ &$\hm 0.148\pm 0.004$ & WIYN \\ 
$3055.690$ &$  1.46$ &$-1.077\pm 0.009$ &$ 1.321\pm 0.012$ &$ 0.973\pm 0.010$ &$-0.003\pm 0.004$ &SMARTS \\ 
$3055.753$ &$  1.09$ &$-1.080\pm 0.004$ &$ 1.314\pm 0.008$ &$ 0.950\pm 0.007$ &$\hm 0.068\pm 0.004$ &MAGELLAN \\ 
$3063.677$ &$  1.47$ &$-1.082\pm 0.008$ &$ 1.362\pm 0.011$ &$ 0.968\pm 0.009$ &$\hm 0.004\pm 0.004$ &SMARTS \\ 
$3064.824$ &$  1.55$ &$-1.077\pm 0.005$ &$ 1.235\pm 0.008$ &$ 0.900\pm 0.007$ &$\hm 0.066\pm 0.004$ &  MDM \\ 
$3080.640$ &$  1.90$ &$-1.083\pm 0.008$ &$ 1.374\pm 0.011$ &$ 0.993\pm 0.009$ &$\hm 0.005\pm 0.004$ &SMARTS \\ 
$3091.573$ &$  1.14$ &$-1.078\pm 0.009$ &$ 1.309\pm 0.012$ &$ 0.938\pm 0.010$ &$-0.006\pm 0.004$ &SMARTS \\ 
$3101.647$ &$  1.11$ &$-1.102\pm 0.011$ &$ 1.377\pm 0.017$ &$ 0.918\pm 0.012$ &$-0.041\pm 0.005$ &SMARTS \\ 
$3104.646$ &$  2.85$ &$-1.105\pm 0.004$ &$ 1.297\pm 0.007$ &$ 0.917\pm 0.007$ &$\hm 0.172\pm 0.004$ & WIYN \\ 
$3108.540$ &$  3.46$ &$-1.119\pm 0.008$ &$ 1.289\pm 0.012$ &$ 0.964\pm 0.010$ &$\hm 0.000\pm 0.004$ &SMARTS \\ 
$3116.582$ &$  1.02$ &$-1.119\pm 0.009$ &$ 1.337\pm 0.013$ &$ 0.953\pm 0.010$ &$-0.002\pm 0.004$ &SMARTS \\ 
$3132.534$ &$  1.07$ &$-1.107\pm 0.009$ &$ 1.300\pm 0.012$ &$ 0.961\pm 0.010$ &$-0.021\pm 0.005$ &SMARTS \\ 
$3136.717$ &$  5.84$ &$-1.132\pm 0.005$ &$ 1.255\pm 0.008$ &$ 0.893\pm 0.007$ &$\hm 0.062\pm 0.004$ &  MDM \\ 
$3138.476$ &$  1.34$ &$-1.093\pm 0.009$ &$ 1.322\pm 0.013$ &$ 1.004\pm 0.011$ &$-0.016\pm 0.005$ &SMARTS \\ 
$3359.787$ &$  2.12$ &$-0.929\pm 0.011$ &$ 1.350\pm 0.014$ &$ 1.085\pm 0.011$ &$-0.006\pm 0.004$ &SMARTS \\ 
$3368.759$ &$  0.66$ &$-0.888\pm 0.023$ &$ 1.445\pm 0.034$ &$ 1.251\pm 0.026$ &$-0.021\pm 0.005$ &SMARTS \\ 
$3393.988$ &$  3.54$ &$-0.867\pm 0.009$ &$ 1.690\pm 0.012$ &$ 1.245\pm 0.009$ &$\hm 0.032\pm 0.004$ &  MDM \\ 
$3394.754$ &$  0.80$ &$-0.870\pm 0.015$ &$ 1.531\pm 0.019$ &$ 1.117\pm 0.031$ &$-0.040\pm 0.005$ &SMARTS \\ 
$3403.785$ &$  1.00$ &$-0.850\pm 0.011$ &$ 1.555\pm 0.014$ &$ 1.199\pm 0.011$ &$-0.011\pm 0.004$ &SMARTS \\ 
$3413.792$ &$  2.84$ &$-0.850\pm 0.009$ &$ 1.694\pm 0.014$ &$ 1.286\pm 0.011$ &$\hm 0.001\pm 0.004$ &SMARTS \\ 
$3417.760$ &$  1.55$ &$-0.832\pm 0.009$ &$ 1.637\pm 0.013$ &$ 1.307\pm 0.015$ &$-0.019\pm 0.004$ &SMARTS \\ 
$3424.727$ &$  0.60$ &$-0.818\pm 0.022$ &$ 1.672\pm 0.035$ &$ 1.216\pm 0.038$ &$-0.026\pm 0.005$ &SMARTS \\ 
$3428.737$ &$  1.19$ &$-0.791\pm 0.014$ &$ 1.709\pm 0.024$ &$ 1.354\pm 0.031$ &$-0.042\pm 0.005$ &SMARTS \\ 
$3431.786$ &$  1.51$ &$-0.836\pm 0.014$ &$ 1.772\pm 0.020$ &$ 1.282\pm 0.014$ &$-0.030\pm 0.005$ &SMARTS \\ 
$3433.726$ &$  1.44$ &$-0.832\pm 0.010$ &$ 1.689\pm 0.015$ &$ 1.230\pm 0.011$ &$-0.009\pm 0.004$ &SMARTS \\ 
$3435.754$ &$  1.71$ &$-0.830\pm 0.008$ &$ 1.634\pm 0.013$ &$ 1.224\pm 0.010$ &$\hm 0.006\pm 0.004$ &SMARTS \\ 
$3442.691$ &$  2.05$ &$-0.833\pm 0.011$ &$ 1.607\pm 0.015$ &$ 1.245\pm 0.011$ &$-0.010\pm 0.004$ &SMARTS \\ 
$3445.734$ &$  2.31$ &$-0.845\pm 0.009$ &$ 1.696\pm 0.015$ &$ 1.254\pm 0.011$ &$-0.005\pm 0.004$ &SMARTS \\ 
$3447.633$ &$  0.82$ &$-0.833\pm 0.013$ &$ 1.630\pm 0.017$ &$ 1.199\pm 0.012$ &$-0.018\pm 0.005$ &SMARTS \\ 
$3449.741$ &$  1.39$ &$-0.823\pm 0.010$ &$ 1.630\pm 0.014$ &$ 1.232\pm 0.011$ &$-0.013\pm 0.004$ &SMARTS \\ 
$3458.682$ &$  0.96$ &$-0.819\pm 0.012$ &$ 1.675\pm 0.017$ &$ 1.264\pm 0.013$ &$-0.032\pm 0.005$ &SMARTS \\ 
$3459.599$ &$  0.81$ &$-0.818\pm 0.015$ &$ 1.604\pm 0.020$ &$ 1.244\pm 0.014$ &$-0.030\pm 0.005$ &SMARTS \\ 
$3461.676$ &$  0.86$ &$-0.825\pm 0.012$ &$ 1.641\pm 0.016$ &$ 1.211\pm 0.012$ &$-0.010\pm 0.005$ &SMARTS \\ 
$3468.600$ &$  2.00$ &$-0.828\pm 0.009$ &$ 1.630\pm 0.014$ &$ 1.253\pm 0.011$ &$-0.001\pm 0.004$ &SMARTS \\ 
$3470.608$ &$  2.68$ &$-0.837\pm 0.011$ &$ 1.543\pm 0.015$ &$ 1.324\pm 0.012$ &$-0.006\pm 0.004$ &SMARTS \\ 
$3478.556$ &$  0.83$ &$-0.810\pm 0.014$ &$ 1.595\pm 0.018$ &$ 1.214\pm 0.013$ &$-0.033\pm 0.005$ &SMARTS \\ 
$3483.539$ &$  1.05$ &$-0.793\pm 0.017$ &$ 1.603\pm 0.026$ &$ 1.677\pm 0.030$ &$-0.037\pm 0.005$ &SMARTS \\ 
$3490.600$ &$  1.49$ &$-0.822\pm 0.011$ &$ 1.653\pm 0.015$ &$ 1.256\pm 0.011$ &$-0.013\pm 0.004$ &SMARTS \\ 
$3491.594$ &$  1.34$ &$-0.808\pm 0.011$ &$ 1.688\pm 0.015$ &$ 1.263\pm 0.011$ &$-0.011\pm 0.004$ &SMARTS \\ 
$3523.497$ &$  0.97$ &$-0.754\pm 0.012$ &$ 1.703\pm 0.017$ &$ 1.270\pm 0.012$ &$-0.017\pm 0.005$ &SMARTS \\ 
$3527.474$ &$  1.20$ &$-0.761\pm 0.011$ &$ 1.753\pm 0.016$ &$ 1.318\pm 0.012$ &$-0.014\pm 0.005$ &SMARTS \\ 
$3718.838$ &$  0.92$ &$-0.787\pm 0.015$ &$ 1.562\pm 0.018$ &$ 1.174\pm 0.013$ &$-0.024\pm 0.005$ &SMARTS \\ 
$3737.018$ &$  4.71$ &$-0.885\pm 0.004$ &$ 1.434\pm 0.007$ &$ 1.127\pm 0.007$ &$\hm 0.099\pm 0.004$ &  MDM \\ 
$3750.790$ &$  0.68$ &$-0.795\pm 0.022$ &$ 1.663\pm 0.035$ &$ 1.331\pm 0.026$ &$-0.030\pm 0.005$ &SMARTS \\ 
$3755.748$ &$  0.63$ &$-0.784\pm 0.023$ &$ 1.613\pm 0.037$ &$ 1.186\pm 0.024$ &$-0.020\pm 0.005$ &SMARTS \\ 
$3761.765$ &$  1.09$ &$-0.820\pm 0.010$ &$ 1.579\pm 0.013$ &$ 1.183\pm 0.010$ &$\hm 0.005\pm 0.004$ &SMARTS \\ 
$3772.763$ &$  1.71$ &$-0.828\pm 0.009$ &$ 1.547\pm 0.013$ &$ 1.175\pm 0.010$ &$\hm 0.007\pm 0.004$ &SMARTS \\ 
$3792.766$ &$  2.47$ &$-0.845\pm 0.011$ &$ 1.706\pm 0.017$ &$ 1.240\pm 0.012$ &$-0.006\pm 0.004$ &SMARTS \\ 
$3815.678$ &$  1.48$ &$-0.851\pm 0.010$ &$ 1.721\pm 0.017$ &$ 1.116\pm 0.012$ &$-0.012\pm 0.005$ &SMARTS \\ 
$3823.631$ &$  2.25$ &$-0.885\pm 0.013$ &$ 1.646\pm 0.018$ &$ 1.228\pm 0.013$ &$-0.016\pm 0.005$ &SMARTS \\ 
$3855.526$ &$  0.59$ &$-0.883\pm 0.030$ &$ 1.522\pm 0.058$ &$ 0.996\pm 0.051$ &$\hm 0.023\pm 0.006$ &SMARTS \\ 
$3884.450$ &$  1.00$ &$-0.908\pm 0.019$ &$ 1.502\pm 0.024$ &$ 1.118\pm 0.024$ &$-0.033\pm 0.005$ &SMARTS \\ 
$3893.475$ &$  0.64$ &$-0.921\pm 0.025$ &$ 1.655\pm 0.035$ &$ 1.130\pm 0.022$ &$-0.032\pm 0.005$ &SMARTS \\ 
$3900.490$ &$  0.94$ &$-0.919\pm 0.012$ &$ 1.548\pm 0.017$ &$ 1.159\pm 0.012$ &$-0.023\pm 0.005$ &SMARTS \\

\enddata
\tablecomments{HJD is the Heliocentric Julian Day --2450000 days.
The goodness of fit of the image, $\chi^2/N_{dof}$, is used to rescale the
formal uncertainties by a factor of $(\chi^2/N_{dof})^{1/2}$.  The Image
columns give the magnitudes of the quasar images relative to the
comparison stars, and the Images A1+A2 column is the sum of the flux from both images.
The $\langle\hbox{Stars}\rangle$ column gives the
mean magnitude of the standard stars for that epoch relative to
their mean for all epochs. 
}
\label{tab:lightcurves}
\end{deluxetable}

\end{document}